\documentclass[
    ,final            
  ]
  {aipproc}

\layoutstyle{6x9}

\def\half{{1\over 2}}

\def\fourth{{1\over4}}

\def\Z{{\mathchoice {\hbox{$\sf\textstyle Z\kern-0.4em Z$}}
{\hbox{$\sf\textstyle Z\kern-0.4em Z$}}
{\hbox{$\sf\scriptstyle Z\kern-0.3em Z$}}
{\hbox{$\sf\scriptscriptstyle Z\kern-0.2em Z$}}}}
\def\abs#1{\left| #1\right|}

\def\square{\kern1pt\vbox{\hrule height 1.2pt\hbox{\vrule width 1.2pt
   \hskip 3pt\vbox{\vskip 6pt}\hskip 3pt\vrule width 0.6pt}
   \hrule height 0.6pt}\kern1pt}
      \def\boxop{{\raise-.25ex\hbox{\square}}}

\def\mn{{\mu\nu}}

\def\e{\,{\rm e}}

\newcommand{\be}{\begin{equation}}
\newcommand{\ee}{\end{equation}\noindent}
\newcommand{\bear}{\begin{eqnarray}}
\newcommand{\ear}{\end{eqnarray}\noindent}
\newcommand{\benn}{\begin{enumerate}}
\newcommand{\enn}{\end{enumerate}}

\def\slash#1{#1\!\!\!\raise.15ex\hbox {/}}
\newcommand{\slD}{\,\raise.15ex\hbox{$/$}\kern-.27em\hbox{$\!\!\!D$}}
\newcommand{\slpartial}{\raise.15ex\hbox{$/$}\kern-.57em\hbox{$\partial$}}
\def\4piTD{{(4\pi T)}^{-{D\over 2}}}
\def\4piT4{{(4\pi T)}^{-2}}

\def\Tintm4{{\dps\int_{0}^{\infty}}{dT\over T}\,e^{-m^2T}
    {(4\pi T)}^{-2}}
\def\Tintm{{\dps\int_{0}^{\infty}}{dT\over T}\,e^{-m^2T}}

\newcommand{\slG}{{{\dot G}\!\!\!\! \raise.15ex\hbox {/}}}

\def\GBd12{{\dot G}_{B12}}

\newcommand{\no}{\noindent}
\def\non{\nonumber}
\def\dps{\displaystyle}

\begin{document}

\title{QED in the worldline representation}


\author{Christian Schubert}{
  address={Instituto de F\'\i{s}ica y Matem\'aticas,
Universidad Michoacana de San Nicol\'as de Hidalgo, \\
Edificio C-3, Ciudad Universitaria, 
C.P. 58040 Morelia, Michoacan, Mexico,\\
schubert@ifm.umich.mx}
}

\begin{abstract}
Simultaneously with inventing the modern relativistic formalism of quantum electrodynamics,
Feynman presented also a first-quantized representation of QED in terms of 
worldline path integrals.  Although this alternative formulation has been studied over the years  
by many authors, only during the last fifteen years
it has acquired some popularity as a computational tool.
I will shortly review here three 
very different techniques which have been developed during the last few years for the
evaluation of worldline path integrals, namely (i) the ``string-inspired formalism'', based
on the use of worldline Green functions, (ii) the numerical ``worldline Monte Carlo formalism'',
and (iii) the semiclassical ``worldline instanton'' approach. 
\end{abstract}

\maketitle


\section{1. Feynman's worldline representation of QED}

\begin{figure}
  \includegraphics[height=.4\textheight]{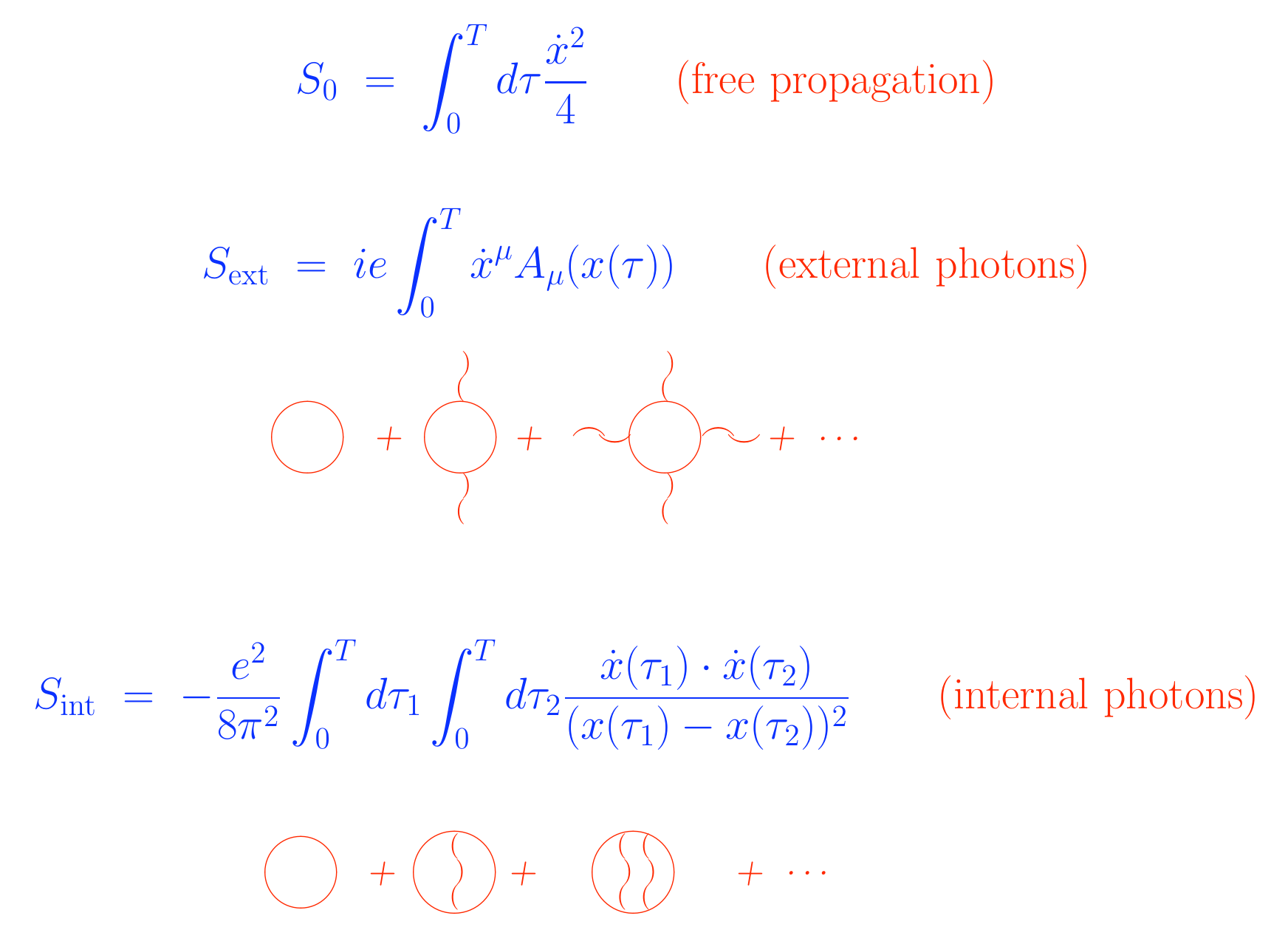}
  \caption{Perturbative expansion of the worldline path integral.}
  \label{wlexpansion}
\end{figure}

In 1950 Feynman presented, in an appendix to one of his groundbreaking papers on the 
modern, manifestly relativistic formalism of perturbative QED \cite{feynman1950}, also
a first-quantized formulation of scalar QED, ``for its own interest as an alternative
to the formulation of second quantization''. There he provides a simple rule for constructing
the complete scalar QED S-matrix by representing virtual scalars and photons in terms
of relativistic particle path integrals, and coupling them in all possible ways. 
Restricting ourselves, for simplicity, to the purely photonic part of the S-matrix (no
external scalars), and moreover to the ``quenched'' contribution (only one virtual scalar), 
this ``worldline representation'' can be given most compactly in terms of
the (quenched) effective action $\Gamma [A]$:

\bear
\Gamma_{\rm scalar}[A] &=&
\int d^4x\, {\cal L}_{\rm scalar}[A] =
\int_0^{\infty}{dT\over T}\,{\rm e}^{-m^2T}
{\displaystyle \int_{x(T)=x(0)}}{\cal D}x(\tau)
\, e^{-S[x(\tau)]}
\label{Gammascal}
\ear
Here $T$ denotes the proper-time of the scalar particle in the loop,
$m$ its mass, and $ \int_{x(T)=x(0)}{\cal D}x(\tau)$ a path integral over
all closed loops in spacetime with fixed periodicity in the proper-time.
The worldline action $S[x(\tau)]$ has three parts,  
 
\bear
S&=& S_0 + S_{\rm ext} + S_{\rm int}\label{Ssplit}
\ear
(see fig. \ref{wlexpansion}).
Of these, the kinetic term $S_0$ describes the free propagation of the scalar, $S_{\rm ext}$
its interaction with the external field, and $S_{\rm int}$ the corrections due to
internal photon exchanges in the loop.
The connection to a standard Feynman diagrammatic description is made simply
by expanding out the two interaction exponentials.

The generalization of this representation to include multiple
scalar loops and open scalar lines is straightforward. It yields
a first-quantized representation of the full
effective action, and thus, by Fourier transformation, of the S-matrix.

While for scalar QED this representation is essentially unique,
when generalizing it to spinor QED one has a number of choices.
First, worldline representations of spin half particles can be derived either
from standard first-order Dirac theory, or from its
second-order formulation (see \cite{morgan} and refs. therein), 
based on the identity

\bear
 ({\slash p}+e\slash A )^2 &=&
 -(\partial + ie A)^2 -{i\over 2}e\, \sigma^\mn F_{\mn}
 \label{idsquare}
 \ear
Contrary to the situation in second-quantized field theory, in the worldline
formalism the second-order approach is the more standard one,
and I will restrict myself to it in this review (see \cite{migdal,fosava} 
for the first-order approach). Using (\ref{idsquare}) one arrives at a
worldline representation for the fermion QED effective action 
$\Gamma_{\rm spinor}[A]$ which, at the one-loop level, differs from
the scalar one (\ref{Gammascal}) only by the addition of a global
factor of $-\half$, and the insertion of a {\it spin factor} $S[x,A]$ under
the path integral \cite{feynman1951},

\bear
S[x,A] &=& {\rm tr}_{\Gamma} {\cal P}
\exp\biggl[{{i\over 2}e\,\sigma^{\mu\nu}
\int_0^Td\tau \, F_{\mu\nu}(x(\tau))}\biggr]
\label{defspinfactor}
\ear
Here ${\rm tr}_{\Gamma}$ denotes the Dirac trace and $\cal P$
path ordering.

A more modern way of writing the same spin factor is 
in terms of an additional Grassmann path integral
\cite{fradkin,bermar,bdzdh,bacalu,bdh},

\bear
S[x,A]  &=&
\int {\cal D}\psi(\tau)
\,
\exp 
\Biggl\lbrack
-\int_0^Td\tau
\Biggl(
\half \psi\cdot \dot\psi -ie \psi^{\mu}F_{\mn}\psi^{\nu}
\Biggr)
\Biggr\rbrack
\label{spinfactorgrass}
\ear
Here the path integration is over the space of anticommuting
functions antiperiodic in proper-time,
$\psi^{\mu}(\tau_1)\psi^{\nu}(\tau_2) = - \psi^{\nu}(\tau_2)\psi^{\mu}(\tau_1)$,
$\psi^{\mu}(T) = - \psi^{\mu}(0)$.
The main advantage of introducing this second path integral is that, as it turns out,
there is a ``worldline'' supersymmetry between the coordinate function
$x(\tau)$ and the spin function $\psi(\tau)$ \cite{bdzdh}. Although this supersymmetry
is broken by the different periodicity conditions for $x$ and $\psi$, it still has
a number of useful computational consequences. In particular, introducing a
worldline superformalism allows one to combine the two path integrals,
and to write down a formula for $\Gamma_{\rm spinor}[A]$ which is completely
analogous to (\ref{Gammascal}), (\ref{Ssplit}), including the internal photon
corrections \cite{ss3}. 

A third, even more subtle, way of implementing spin on the worldline is the "Polyakov spin factor",
a purely geometric quantity depending only on the worldline 
itself \cite{strominger,polyakov}. This aspect of spin has been studied by a number of
authors, but usually using the first-order formalism; the form of the spin factor
appropriate for the second-order formalism has been established only
recently \cite{giehae}.

A vast amount of work has been done on these QED worldline representations
and their generalizations to other background fields and couplings 
(see \cite{report} for an extensive list of references).
Nevertheless, most of the earlier work on this subject is concerned with formal aspects of relativistic particle Lagrangians, rather than with attempts at performing state-of-the-art calculations in
quantum field theory (notable exceptions are \cite{halsie,hajese} and \cite{afalma}). 
It is only during the last fifteen years that the usefulness
of the first-quantized approach as an alternative to standard Feynman diagrammatic methods
has been seriously investigated. This development was triggered by string theory, where
first-quantized path integrals are the standard tool for perturbative calculations of scattering
amplitudes, and by the fact that many amplitudes in quantum field theory can be represented
as infinite string tension limits of the corresponding string amplitudes. Bern and Kosower \cite{berkos}
investigated this limit in detail for the case of the QCD $N$ gluon amplitudes, and found in
this way a new set of rules for the construction of these amplitudes. Shortly later, Strassler's
work \cite{strassler} showed that the same type of ``string-inspired'' representation of photon/gluon 
amplitudes can also be obtained more directly by evaluating the corresponding first-quantized path integrals using appropriate worldline Green functions. This approach was then generalized 
to some cases of multiloop amplitudes \cite{ss2,ss3,rolsat,dashsu}, as well as to QED amplitudes in a constant external field \cite{shaisultanov,rescsc} (see \cite{report} for a review).  
Moreover, the success of this program led to the development of
alternative techniques for the calculation of worldline path integrals. In this talk,
I will restrict myself to the computational aspects of the worldline representation, and to the
prototypical case, QED. I will discuss, in turn, three quite different methods which are
now available for the computation of the scalar/spinor QED worldline path integrals,
(i) the original ``string-inspired'' approach, 
(ii) the numerical ``worldline Monte Carlo'' method and
(iii) the semiclassical ``worldline instanton'' technique.

\section{2. The ``string-inspired'' approach}
\label{stringinspired}

The strategy in the ``string-inspired approach'' is simple. The path integral(s) will be
manipulated into Gaussian form, after which they can be performed using worldline
correlators with the appropriate periodicity properties.
For the one-loop path integrals (\ref{Gammascal}),(\ref{spinfactorgrass}) those are
\cite{polyakovbook,strassler}

\bear
\langle y^{\mu}(\tau_1)y^{\nu}(\tau_2) \rangle
&=&
-G_B(\tau_1,\tau_2)\, \delta^{\mu\nu} \non\\
G_B(\tau_1,\tau_2) &=& \vert \tau_1 -\tau_2\vert 
-{1\over T}\Bigl(\tau_1 -\tau_2\Bigr)^2\non\\
&&\non\\
\langle \psi^{\mu}(\tau_1)\psi^{\nu}(\tau_2)\rangle
&=&
G_F(\tau_1,\tau_2)\, \delta^{\mu\nu} \non\\
G_F(\tau_1,\tau_2) &=& {\rm sign}(\tau_1 - \tau_2)\non\\
\label{green}
\ear
Here $y^{\mu}(\tau) \equiv x^{\mu}(\tau) - x_0^{\mu}$, where $x_0^{\mu}$ denotes
the loop center of mass, $ x^{\mu}_0 \equiv {1\over T}\int_0^T d\tau\, x^{\mu}(\tau)$.
There is a certain freedom in choosing the coordinate correlator $G_B$ (see, e.g., \cite{report}); 
the one given above is usually the most convenient one, but some of the references given below
use the alternative $G_B(\tau_1,\tau_2)=\abs{\tau_1-\tau_2}-(\tau_1+\tau_2) + {2\over T}\tau_1\tau_2$.

\no
Now, to get gaussian form,

\begin{itemize}

\item
Expand all the interaction exponentials $ \e^{-S_{\rm ext}[x(\tau)]}$ etc.

\item
For the  effective action: 
Taylor expand the external field at the loop center of mass, 

\bear
A^{\mu}(x(\tau)) = \e^{y(\tau)\cdot\partial}A^{\mu}(x_0)
\label{expA}
\ear

\item
For the  $N$ photon amplitudes :
Expand the field in $N$ plane waves,

\bear
A^{\mu}(x(\tau)) = \sum_{i=1}^N \,\varepsilon_i^{\mu}\e^{ik_i\cdot x(\tau)}
\label{planewave}
\ear

\item
Exponentiate the denominator of the photon insertion terms,

\bear
-{e^2\over 8\pi^2}\int_0^Td\tau_1\int_0^Td\tau_2 {\dot x(\tau_1)\cdot\dot x(\tau_2)\over
(x(\tau_1)-x(\tau_2))^2}
&=& \non\\
&&\non\\
&&\hspace{-260pt} 
-{e^2\over 2}
\int_0^{\infty}
{d\bar T \over
(4\pi\bar T)^2}
\int_0^Td\tau_1
\int_0^Td\tau_2 \,\,\dot x(\tau_1)\cdot\dot x(\tau_2)
\,{\rm exp}\, \Biggl\lbrack
-{(x(\tau_1)-x(\tau_2))^2\over 4\bar T}
\Biggr\rbrack\non\\
\label{expinsert}
\ear

\end{itemize}
For example, for the four-photon amplitude in scalar QED this procedure yields
the following integral representation:

\bear
\Gamma
[k_1,\varepsilon_1;\ldots ;k_4,\varepsilon_4]
&=&
e^4
{\dps\int_{0}^{\infty}}{dT\over T}
{[4\pi T]}^{-2}
e^{-m^2T}
\prod_{i=1}^4 \int_0^T 
d\tau_i\,
Q_4
\e^{G_{Bij} k_i\cdot k_j}\non\\
\label{fourphoton}
\ear
\vspace{-35pt}

\bear
Q_4
&=&
Q_4^4 + Q_4^3 + Q_4^2 - Q_4^{22}
\label{decompQ4}
\ear\no
\vspace{-20pt}
\bear
Q_4^4 &=& 
\dot G_{B12}
\dot G_{B23}
\dot G_{B34}
\dot G_{B41}
Z_4(1234)
+ 2 \,\, {\rm permutations}
\non\\
Q_4^3 &=&
\dot G_{B12}
\dot G_{B23}
\dot G_{B31}
Z_3(123)
\dot G_{B4i}
\varepsilon_4\cdot k_i
+ 3 \,\, {\rm perm.}
\non\\
Q_4^2 &=&
\dot G_{B12}\dot G_{B21}
Z_2(12)
\biggl\lbrace
\dot G_{B3i}
\varepsilon_3\cdot k_i
\dot G_{B4j}
\varepsilon_4\cdot k_j
+\half
\dot G_{B34}
\varepsilon_3\cdot\varepsilon_4
\Bigl[
\dot G_{B3i}
k_3\cdot k_i
-
\dot G_{B4i}
k_4\cdot k_i
\Bigr]
\biggr\rbrace
\non\\
&&
+ \, 5 \,\, {\rm perm.}
\non\\
Q_4^{22} &=&
\dot G_{B12}\dot G_{B21}
Z_2(12)
\dot G_{B34}\dot G_{B43}
Z_2(34)
+ 2 \,\, {\rm perm.}
\non\\
\label{Q4m}
\ear\no
\vspace{-30pt}

\bear
Z_2(ij)&\equiv&
\varepsilon_i\cdot k_j
\varepsilon_j\cdot k_i
-\varepsilon_i\cdot\varepsilon_j
k_i\cdot k_j
\non\\
Z_n(i_1i_2\ldots i_n)&\equiv&
{\rm tr}
\prod_{j=1}^n 
\Bigl[
k_{i_j}\otimes \varepsilon_{i_j}
- \varepsilon_{i_j}\otimes k_{i_j}
\Bigr]
\quad (n\geq 3)
\non\\
\label{Zn}
\ear\vspace{-20pt}

\no
Here each of the integrals $\int_0^T d\tau_i$ represents one of the four photon legs 
moving around the scalar loop. $G_{Bij}$ stands for $G_B(\tau_i,\tau_j)$ and 
$\dot G_{Bij}$ for its derivative. Repeated indices are to be summed
over $1,\dots,N$. In writing down the integrand, a number of 
integrations by parts have already been performed which eliminated all second
derivatives of the $G_{Bij}$'s that initially appear in the factor $Q_4$. Moreover,
this factor has been decomposed into terms $Q_4^m$ according to ``cycle content'', indicated 
by the superscript, where a ``cycle'' is a factor of 
$\dot G_{B_{i_1i_2}}\dot G_{B_{i_2i_3}}\dot G_{B_{i_3i_4}} \cdots \dot G_{B_{i_ni_1}}$.
Such a `$\tau$-cycle' always gets multiplied by a `Lorentz-cycle' $Z_n(i_1i_2\ldots i_n)$,
which is the trace of the products of the field strength tensors associated to the 
corresponding legs. The various $Q_4^m$'s are individually gauge invariant.
This representation can be generalized to an arbitrary number of photons, and is
unique if one requests manifest permutation symmetry in the photon legs 
\cite{berkos,strassler,nphoton}. The generalization to higher loop photon amplitudes
can be achieved either using (\ref{expinsert}), or by explicit sewing of pairs
of external photons, or more directly using 
generalized worldline Green functions adapted to the
multiloop graph topologies \cite{ss2,rolsat,daisie}.

In all cases
the arising multiple integrals are equivalent to standard Feynman parameter integrals.
For example, in (\ref{fourphoton}) and its $N$-point generalizations,
the prefactor $Q_N$ relates to a Feynman numerator and the exponential
factor to the universal denominator of one-loop $N$-point integrals (see, e.g., \cite{itzzub}).
However, the string-inspired representation has a number of 
interesting properties which are not
usually manifest in standard Feynman parameter calculations:

\begin{enumerate}

\item
The use of the cycle notation allows a more compact way of writing the $N$ photon amplitudes
than usual. 

\item
There is a simple ``cycle replacement rule" which allows one to construct the 
integrand of the spinor loop $N$ photon amplitude from the scalar loop one
\cite{berkos,strassler}. This implies that, in this formalism, the calculations of
the same quantity in scalar and spinor QED are not independent; any 
spinor QED calculation yields the corresponding scalar QED result as a
byproduct. 

\item
The integral (\ref{fourphoton}) and its higher $N$ generalizations represent the whole amplitude,
with no need to sum over permutations. This property is not very relevant at the one-loop
level, but becomes interesting at higher loop orders. In the QED case, it generally allows
one to combine into one integral all contributions from Feynman diagrams which can be
identified by letting photon legs slide along scalar/electron loops or lines. As an example,
we show in fig. \ref{3loopphotonprop} the "quenched" contributions to the 
three-loop photon propagator.

\begin{figure}
  \includegraphics[height=.1\textheight]{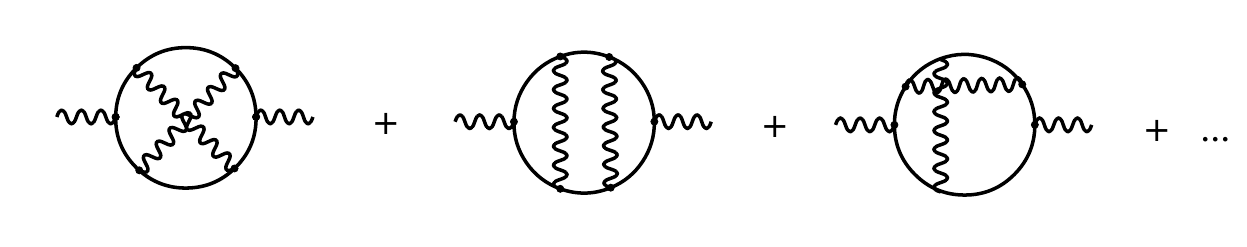}
 \caption{Diagrams contributing to the three loop QED photon propagator.}
 \label{3loopphotonprop}
\end{figure}

This property is particularly interesting in view of the fact that it is precisely 
this type of sums of diagrams which in QED generally leads to extensive cancellations
between diagrams, and to final results which are substantially simpler than intermediate 
ones (see \cite{brdekr} and refs. therein).  
And for the two-loop QED $\beta$ function indeed a way was found 
for calculating the corresponding integral in a way which avoided splitting up the
multiple parameter integral into sectors with a fixed ordering of the photon legs,
and which led to dramatic simplifications \cite{ss3}. However, so far no generalization
of the method used there to higher loop orders has been found.

\item
The string-inspired method provides a 
particularly convenient way of implementing constant external fields in QED calculations.
Photon amplitudes or effective lagrangians in a constant field are obtained from the corresponding vacuum quantities simply by
substituting the vacuum Green functions 
$G_{B,F}(\tau_1,\tau_2)$
by appropriate field-dependent Green functions   
${\cal G}_{B,F}(\tau_1,\tau_2;F)$, and
by a change of the free worldline path integral determinants
\cite{shaisultanov,rescsc} (see also \cite{mckshe}). 
In particular, the "cycle replacement rule" 
carries over to the constant field case.

\end{enumerate}

The efficiency of the technique for
QED in a constant field  has, at the one-loop level, been demonstrated by
recalculations of the photon splitting amplitude in a magnetic field \cite{adlsch}, and
of the one-loop vacuum polarization in a general constant field
\cite{vp1}. At the two-loop level it has been extensively applied 
to the QED effective Lagrangian in a constant field.
This includes recalculations of the standard two-loop Euler-Heisenberg Lagrangians
\cite{rescsc,frss,kors}, closed-form expressions for the weak field expansion coefficients of the
magnetic two-loop Euler-Heisenberg Lagrangian \cite{dhrs}, and
a generalization of these Lagrangians to the case of a self-dual Euclidean field
\cite{sd1}. I will show here only the last result, which is particularly nice:
a Euclidean self-dual field fulfills  
$F_{\mu\nu} = \half \varepsilon_{\mu\nu\alpha\beta}F^{\alpha\beta}$,
which implies that the square of the field strength tensor is proportional to the
unit matrix,  $F_{\mu\nu}F^{\nu\lambda} =  -f^2 \delta^{\lambda}_{\mu}$.
Remarkably, this simplifies matters so much that all parameter integrals
can be done in closed form, leading to the following explicit formulas
for the two-loop scalar and spinor QED effective Lagrangians in a constant
self-dual field \cite{sd1},

\bear
{\cal L}_{\rm scalar}^{(2)}(\kappa)
&=&
\alpha \,{m^4\over (4\pi)^3}\frac{1}{\kappa^2}\left[
{3\over 2}\xi^2 (\kappa)
-\xi'(\kappa)\right]
\nonumber\\
{\cal L}_{\rm spinor}^{(2)}(\kappa)
&=&
-2\alpha \,{m^4\over (4\pi)^3}\frac{1}{\kappa^2}\left[
3\xi^2 (\kappa)
-\xi'(\kappa)\right]
\nonumber\\
\label{ehsd}
\ear\no
Here $ \kappa = \frac{m^2}{2ef}$ and 

\bear
\xi(x)\equiv -x\Bigl({\Gamma'(x)\over\Gamma(x)} -\ln(x)+{1\over 2x}\Bigr)
\label{defxi}
\ear
The simplicity of these results made it possible to use
this self-dual case for studying the generic properties of the weak and strong field
expansions of Euler-Heisenberg Lagrangians \cite{sd2}, as well as the corresponding
maximally helicity violating components of the two-loop $N$-photon amplitudes
in the low energy limit \cite{sd1}.

Apart from the effective action in a constant field, the string-inspired method
has also been extensively applied to the calculation of the full one-loop QED effective action in
an arbitrary background field, using the derivative expansion \cite{ss1,cadhdu,gussho}. 

Moreover, at the one-loop level the method has been generalized to the finite
temperature case, both to obtain representations of the $N$ - photon amplitudes \cite{mckrebthermal,sato} and for the effective action in a general background \cite{shovkovy}. 

Finally, as should be clear from the above, work on the string-inspired technique in QED so far 
has been concerned mainly with purely photonic amplitudes. The extension to amplitudes involving
also external scalars \cite{dashsu} or fermions \cite{mckreb,karkto} is possible without
difficulties, although its practical usefulness is difficult to judge at present
due to a lack of state-of-the-art applications.


\section{3. The ``worldline Monte Carlo'' approach}
\label{montecarlo}

During the last few years it was found that Feynman's worldline
representation (\ref{Gammascal}) and its various generalizations are
also very amenable to a direct numerical evaluation using
standard Monte Carlo techniques \cite{gielan,schsta,gilamo}.
At the one-loop level, this approach has been shown to work well
for QED effective actions in quite
general background fields. This includes singular fields such as a 
magnetic field of the step-function type \cite{gielan}. It also 
extends to the imaginary part of the effective action \cite{giekli}, 
to be discussed in section 4 below.
First steps towards a multiloop extension have been taken in
\cite{gisava}, although, as with all numerical approaches to quantum field
theory, implementing the full renormalization program in such a formalism poses
a formidable challenge.  

The Monte Carlo approach appears to hold particular promise for the calculation
of Casimir energies for arbitrary geometries \cite{gilamo,gieklijpa,giekliprl96}. 
Although this has so far been done only for scalar fields, not for the QED case,
it shall be discussed here since it is interesting to see how 
Dirichlet boundary conditions can be implemented at the level of the
path integral (see the recent \cite{bacola}) for a treatment of boundary conditions
in the ``string-inspired'' approach).   
For a (massless) scalar field in a background potential $V(x)$, 
the gauge coupling term in Feynman's path integral (\ref{Gammascal})
has to be replaced by  $-\int_0^Td\tau\, V(x(\tau))$.
Dirichlet boundary conditions on an (infinitely thin) surface $\Sigma$ can 
be implemented by choosing 

\bear
V({\bf x}) &=& \lambda \int_{\Sigma}d\sigma\, \delta^3({\bf x} - {\bf x_{\sigma}})
\label{VSigma}
\ear 
with $\lambda \to\infty$. 
Considering the case of two disjoint surfaces $\Sigma=\Sigma_1\bigcup\Sigma_2$,
it can then easily be shown \cite{gilamo} that the Casimir interaction energy
between the surfaces is given by
\footnote{When comparing with \cite{gilamo} note that they use a different
normalization of the free path integral.}

\bear
E_{\rm Casimir} &=& -\half {1\over\sqrt{4\pi}} 
\int_0^{\infty}{dT\over T^{3\over 2}}
\int d^3x_0 
{\displaystyle \int}{\cal D}{\bf x}(\tau)
\,e^{-\fourth \int_0^Td\tau\, \dot {\bf x}^2}
\,\Theta_{\Sigma}[{\bf x}(\tau)]
\label{ECasimir}
\ear
where $\Theta_{\Sigma}=1$ if the path $\bf x(\tau)$ intersects both $\Sigma_1$
and $\Sigma_2$, and $\Theta_{\Sigma}=0$ otherwise. The trivial time component of the
path integral has been integrated out. $\bf x_0$ denotes the loop center of mass of the
remaining integral over spatial loops;  the energy density at a point $\bf x_0$ is obtained
by restricting the path integral to worldloops with that point as their common center
of mass. 

\begin{figure}[htb]
\includegraphics[height=.2\textheight]{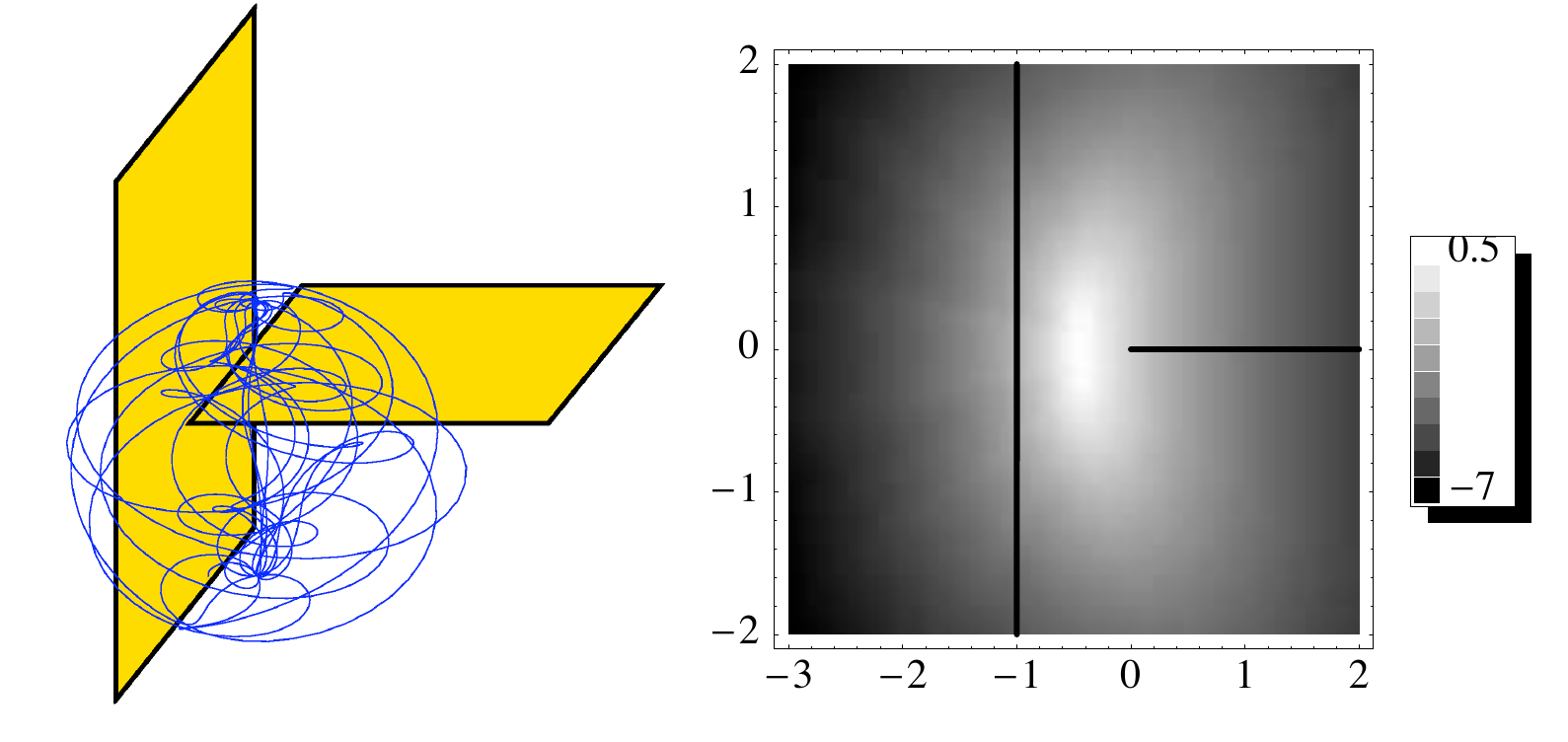}
\caption{Left panel: sketch of the perpendicular-plates configuration with
(an artist's view of) a typical worldline that intersects both plates. Right panel:
Density plot of the effective action density $\cal L$ for the perpendicular plates
case; the plot shows $\ln(2(4\pi a^2)^2\abs{{\cal L}})$. }
\label{figure3}
\end{figure}

Figure \ref{figure3} shows the Casimir energy density 
obtained by a numerical evaluation of (\ref{ECasimir})
for the case of two infinite perpendicular plates at separation $a$
\cite{gieklijpa}
\footnote{Fig. 3 is reprinted from \cite{gieklijpa} 
with the permission of the authors and of the American Institute of
Physics.}

\vspace{25pt}

\begin{figure}[htb]
\includegraphics[height=.25\textheight]{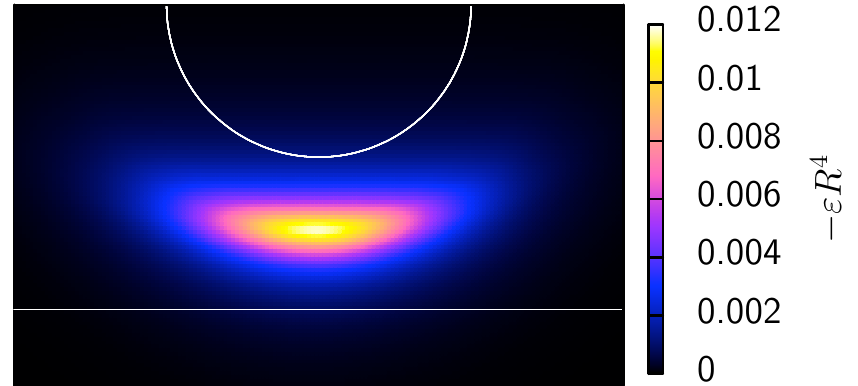}
\caption{Contour plot of the negative Casimir interaction energy density
for a sphere of radius $R$ above an infinite plate; the sphere-plate
separation has been chosen as $a=R$. The plot results from a pointwise evaluation
of eq. (\ref{ECasimir}) using worldlines with a common center of mass.}
\label{figure4}
\end{figure}

\begin{figure}[htb]
\includegraphics[height=.25\textheight]{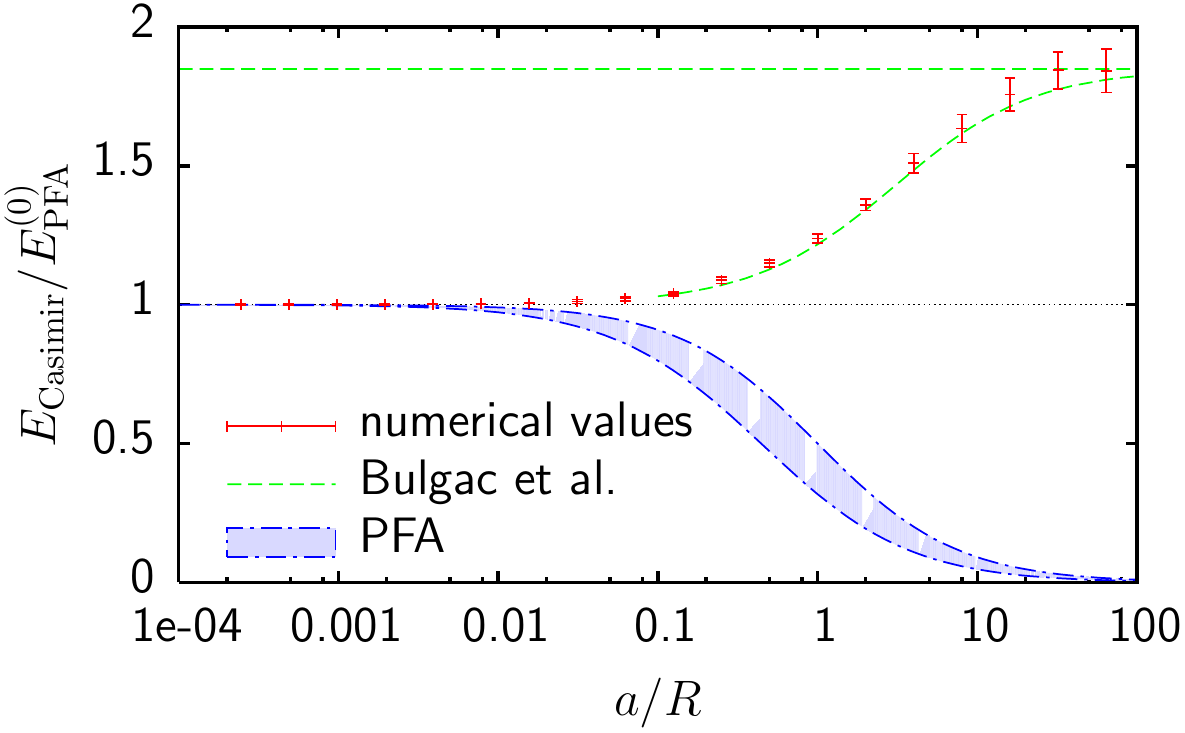}
\caption{Casimir interaction energy of a sphere with radius $R$ and an infinite plate vs. the curvature parameter $a/R$. The energy is normalized with respect to the zeroth-order proximity force approximation (PFA) \cite{giekliprl96}. The numerical worldline result is compared to the
exact result of \cite{bumawi} and the PFA estimate.} 
\label{figure5}
\end{figure}

A particularly nice example is the case of a sphere of radius $R$ and an infinite plate at a distance
$a$ \cite{giekliprl96}. Here an exact result for the Casimir interaction energy is known as a function 
of $a/R$ for 
$a/R {\phantom{x}_> \atop \phantom{x}^\sim} \,\, 0.1$ \cite{bumawi}. 
Fig. \ref{figure4} shows the result
of the worldline evaluation for this case. As can be seen from fig. \ref{figure5}, the
result is in very good agreement with \cite{bumawi} for the whole parameter range
\footnote{Figs. \ref{figure4} and \ref{figure5} are reprinted 
from \cite{giekliprl96} with the permission of the authors and
of the American Physical Society (copyright by APS,  http://link.aps.org/abstract/PRL/v96/e220401)
.}.

\vspace{20pt}

\vspace{20pt}

\section{4. The  ``worldline instanton'' approach}
\label{wlinstanton}

We come now to a
third approach which is more specialized than the two
previous ones, since it applies only to the imaginary part
of amplitudes or the effective action. Although the basic idea
was presented by Affleck et al. in 1982 \cite{afalma}, 
it seems not to have been followed up on until
very recently \cite{wlinst1,wlinst2,dunwan}. 

The work of Affleck et al. concerned the imaginary part of the
scalar QED effective Lagrangian in a constant electric field.
This quantity has been of much interest ever since Schwinger,
building on earlier work by Sauter, Heisenberg, Euler, 
and Weisskopf \cite{sauter,eulhei,weisskopf},
showed that its existence implies the
possibility of electron-positron pair creation in vacuum by the
electric field  \cite{schwinger51}.  
For small production rates this rate is
simply given by twice the imaginary part itself,
$P_{\rm production} \approx 2{\rm Im}{\cal L}[E]$.
Moreover, Schwinger was able to explicitly
calculate the imaginary part in terms of a sum
of exponentials,

\bear
{\rm Im}\, {\cal L}_{\rm spinor}(E) &=&  \frac{m^4}{8\pi^3}
\beta^2\, \sum_{n=1}^\infty \frac{1}{n^2}
\,\exp\left[-\frac{\pi n}{\beta}\right]
\label{ImSpin}
\ear
 ($ \beta = eE/ m^2$). In this sum the
$n$th term relates to the
coherent production of $n$ pairs by the field.
The appearance of $E$ in the denominator of the exponents 
indicates that the pair creation effect is nonperturbative in
nature. It also suggests an interpretation as a tunnel 
effect where a virtual electron-positron pair 
separates out along the field lines 
and extracts a sufficient amount of energy from the field
to turn real.

The corresponding formula for scalar QED differs only by a global factor
and signs (due to the difference in statistics),

\bear
{\rm Im} {\cal L}_{\rm scalar}(E) &=& - \frac{m^4}{16\pi^3}
\beta^2\, \sum_{n=1}^\infty \frac{(-1)^n}{n^2}
\,\exp\left[-\frac{\pi n}{\beta}\right]
\label{ImScal}
\ear
The production rates are exponentially small for
 
\bear
E \ll E_{\rm crit} =  {m^2\over e}\, = 1.3 \times 10^{18}\, {\rm V/m}
\label{Ecrit}
\ear
Until a few years ago producing an electric field close to this critical field strength 
$E_{\rm crit}$, and with a sufficient
spatial extension, appeared far out of the reach of laboratory experiments.
However, due to recent advances in laser technology it seems now
conceivable that pair production could be seen in laser fields in the near
future. The optical laser POLARIS, under construction at the Jena high-intensity laser facility,
is projected to reach a maximal field strength of 
$ E_{\rm max} \approx 2 \times 10^{14}\, {\rm V/m}$
\cite{polaris}, while the European 
X-ray free electron laser (XFEL), under construction at DESY,
is expected to come even closer,
$ E_{\rm max} \approx 1.2 \times 10^{16}\, {\rm V/m} $
\cite{tesla}.

For realistic laser experiments it is usually far from clear whether the constant
field approximation is justified; it would be preferable to have generalizations of Schwinger's formula 
(\ref{ImSpin}) to inhomogeneous and time-dependent fields. 
This is a subject which has been pursued by
many authors, and many results have been obtained over the years (see, e.g.
\cite{keldysh,breitz,nikishov,narnik,popov,stone,grmurabook,basefr,dunhal,kimpag}). 
However, considering this large volume of work there is a surprising lack of
variety in the calculation methods used; apart from a few special field configurations, 
virtually all results of a more general nature have been obtained 
by WKB, or some variant of it. The formalism developed below is, while similar in spirit
to WKB, technically quite different, and apparently more general.

Let us start with retracing Affleck et al.'s \cite{afalma}
recalculation of the Schwinger formula for scalar QED, (\ref{ImScal}).  
At one loop, Feynman's representation (\ref{Gammascal}) reads

\bear
\Gamma_{\rm scalar} [A] &=&
\int_0^{\infty}{dT\over T}\, \e^{-{m^2}T}
\int {\cal D}x 
\, \e^{-\int_0^Td\tau 
\bigl({\dot x^2\over 4} +ieA\cdot \dot x \bigr)}
\label{Gammascaloneloop}
\ear
Rescaling $\tau = Tu$,  this becomes

\bear
\Gamma_{\rm scalar} [A] &=&
\int_0^{\infty}{dT\over T}\, \e^{-{m^2}T}
\int {\cal D}x 
\, \e^{-\Bigl({1\over T}\int_0^1du \,
\dot x^2 +ie\int_0^1du A\cdot \dot x 
\Bigr) }
\label{rescale}
\ear
The $ T$  integral has a
stationary point at 

\bear
T_c &=& {\sqrt{\int du\,\dot x^2}\over m}
\label{Tcrit}
\ear
This allows us to calculate its imaginary
part using a stationary phase approximation,
yielding

\bear
{\rm Im} \Gamma_{\rm scalar} &=&
{1\over m}
\,{\rm Im} \int {\cal D}x \, 
\sqrt{2\pi\over T_c}
\e^{-\Bigl(m\sqrt{\int du \,\dot x^2} 
+ie\int_0^1 du A\cdot \dot x
\Bigr)}
\label{spapprox}
\ear
with a  new worldline action,

\bear
S &=& m\sqrt{\int du\, \dot x^2} + ie \int_0^1duA\cdot \dot x
\label{newaction}
\ear
We would like to calculate the remaining path integral
using a stationary phase approximation, too.
The action (\ref{newaction}) is stationary if

\bear
m{\ddot x_{\mu}\over \sqrt{\int du\,\dot x^2}} &=& ie F_{\mn}\dot x_{\nu}
\label{condstat}
\ear
Contracting with $\dot x^{\mu}$ yields 
$\dot{x}^2={\rm constant}\equiv a^2$, 
and thus 
 
\bear
m\ddot x_{\mu} = iea F_{\mn}\dot x_{\nu}
\label{Lorentz}
\ear
Thus the extremal action trajectory $ x^{\rm cl}(u)$, the
``worldline instanton'', 
will be a periodic solution of the Lorentz force equation,
with a parameter $a$ to be determined by the
periodicity condition. Once the instanton is found,
its worldline action immediately provides a 
semiclassical approximation 
for the imaginary part of the effective Lagrangian.
Closer inspection shows that this approximation
also corresponds to a weak field approximation \cite{wlinst1}:

\bear
{\rm Im}{\cal L}_{\rm scalar}(E)  
\,\,
\stackrel{\rm E\to 0}{\sim}
\,\, {\rm e}^{-S[x^{\rm cl}]}\label{scapprox}
\ear
For a the case of a constant electric field, 
${\vec E} = (0,0,E) = const.$,
it is easy to see that the periodicity condition is solved by

\bear
a_n &=& {m\over eE} \,2n\pi, \qquad n \in {\bf Z}^+
\label{an}
\ear
The worldline instantons are simply circles in the $t-z$ plane, with a winding number $n$:

\bear
x_n^{\rm cl}(u) &=& {m\over eE}\,\Bigl(x_1,x_2,{\rm cos}(2n\pi u),{\rm sin}(2n\pi u)\Bigr)\non\\
S[x_n^{\rm cl}] &=& n\pi\, {m^2\over eE}\non\\
\label{standardinstanton}
\ear
The evaluation of the worldline action on the $n$th instanton thus yields just the
$n$th exponent in Schwinger's formula (\ref{ImScal}). Moreover, Affleck et al.
were able to compute also the prefactor, which here involves the determinant
of fluctuations around the instanton path (see below). Thus, this approach provides
a very simple and elegant rederivation of Schwinger's result for scalar QED.

We will now sketch how to generalize this approach to general electric fields,
as well as to the spinor QED case; see \cite{wlinst1,wlinst2,dunwan} for the details.
Starting all over from  Feynman's representation of the one-loop effective action in scalar QED
(\ref{Gammascal}), for the general case we prefer not to eliminate the
$T$ integral, but rather to first seek a stationary phase approximation for the
path integral. The stationarity condition is again the Lorentz force equation,

\bear
\ddot x_{\mu} &=& 2ie F_{\mn}(x) \dot x_{\nu}
\label{lorentz2}
\ear
We fix a point on the loop:

\bear
{\displaystyle \int_{x(T)=x(0)}}\hspace{-30pt}{\cal D}x(\tau)
\, e^{-S[x(\tau)]}
=
\int d^4x^{(0)}
{\displaystyle \int_{x(T)=x(0)=x^{(0)}}}
\hspace{-50pt}{\cal D}x(\tau)
\, e^{-S[x(\tau)]}
\label{fixapoint}
\ear
Let us assume that we have found a 
worldline instanton $x^{\rm cl}(\tau)$, a classical solution with
 $x(T)=x(0)=x^{(0)}$. We expand around $ x^{\rm cl}$:
  
 \bear
 x_{\mu}(\tau) &=& x^{\rm cl}_{\mu}(\tau) + \eta_{\mu}(\tau), \qquad
 \eta_{\mu}(0) = \eta_{\mu}(T) = 0.
 \label{fluct}
 \ear
 Obtain the  operator of quadratic fluctuations 
 (Jacobi or Hessian matrix) $\Lambda_{\mn}$
 
 \bear
 \Lambda_{\mn} = - \half \delta_{\mn} {d^2\over d\tau^2}
 - {d\over d\tau}Q_{\nu\mu} + Q_{\mn}{d\over d\tau}
 +R_{\mn},
 \label{defLambda}
 \ear
  where
 
 \bear
 Q_{\mn} &=& {\partial^2 L\over \partial x_{\mu} \partial \dot x_{\nu}},
 \qquad
 R_{\mn} = {\partial^2 L\over \partial x_{\mu} \partial  x_{\nu}} \label{defQR}
 \ear
  Find the  zero modes  $\eta_{\nu}^{(\lambda)}(\tau)$ 
 of  $\Lambda_{\mn}$,
 
 \bear
 \Lambda_{\mu\nu} \eta_{\nu}^{(\lambda)} = 0
 \label{Lambdazeromodes}
 \ear
  with initial value conditions
 
 \bear
 \eta_{\nu}^{(\lambda)} (0) &=& 0, \qquad \dot \eta_{\nu}^{(\lambda)}(0) = \delta_{\nu\lambda},
 \qquad (\mu,\nu = 1,2,3,4)
 \label{condinit}
 \ear
 Evaluate them at $\tau =T$. Then, the final result for the semiclassical approximation  becomes
\cite{wlinst2}

\bear
{\displaystyle \int_{x(T)=x(0)=x^{(0)}}}\hspace{-60pt}{\cal D}x(\tau)
\, e^{-S[x(\tau)]}
&=&
{\e^{i\theta}\e^{-S[x^{\rm cl}](T)}\over (4\pi T)^2}
\sqrt{
\biggl\vert {\rm det}\Bigl[\eta^{(\lambda)}_{\mu,{\rm free}}(T)\Bigr]\biggr\vert\over
\biggl\vert {\rm det}\Bigl[\eta^{(\lambda)}_{\mu}(T)\Bigr]\biggr\vert}
\label{final}
\ear
Note that the calculation of this quite general fluctuation determinant has been reduced to 
finding the determinant
of the $4\times 4$ matrix of zero modes. This remarkable simplification relies on
a theorem by Levit and Smilansky \cite{levsmi}.
Finally, at the very end one does  $\int dT$  using the stationary phase method again.

For a number of classical special cases of ``planar'' non-constant fields, such as the 
single-pulse time dependent field \cite{narnik,popov}, the single-bump
space dependent field \cite{nikishov}, the sinusoidal time-dependence \cite{breitz,popov},
and the sinusoidal space-dependence, the worldline instantons can be found
explicitly in terms of special functions \cite{wlinst1}, leading to simple explicit
formulas for the semiclassical exponent. For example, for the single-pulse field

\bear
E(t)&=&E\, {\rm sech}^2(\omega\,t)
\label{singlepulse}
\ear
the stationary action is \cite{wlinst1}

\vspace{-10pt}

\bear
S_{\rm pulse}&=&n\, \frac{m^2 \pi}{e E}\left(\frac{2}{1+\sqrt{1+\gamma^2}}\right)
\label{Ssinglepulse}
\ear
where $n$ is the winding number and 
$\gamma\equiv\frac{m\omega}{eE}$ the ``adiabaticity  parameter'' \cite{keldysh}.
For the analogous single-bump field

\bear
E(x_3)&=&E\, {\rm sech}^2(kx_3)
\label{singlebump}
\ear
one finds

\vspace{-10pt}

\bear
S_{\rm bump}&=&n\, \frac{m^2 \pi}{e E}\left(\frac{2}{1+\sqrt{1-\tilde\gamma^2}}\right)
\label{Ssinglebump}
\ear
where
$\tilde \gamma =  m k/eE$.
Note that 
$S_{\rm pulse}$  decreases with  $\gamma$, so that the pair production rate
increases, while it is the other way round for $S_{\rm bump}$.
We believe that this is just an instance of a quite general fact: 
{\it Inhomogeneity in time tends to shrink the size of the wordline instantons,
leading to an increase in the pair production rate; 
spatial inhomogeneity increases the instanton size and decreases the pair production rate}.

\begin{figure}[h]
\centerline{\includegraphics[height=.29\textheight]{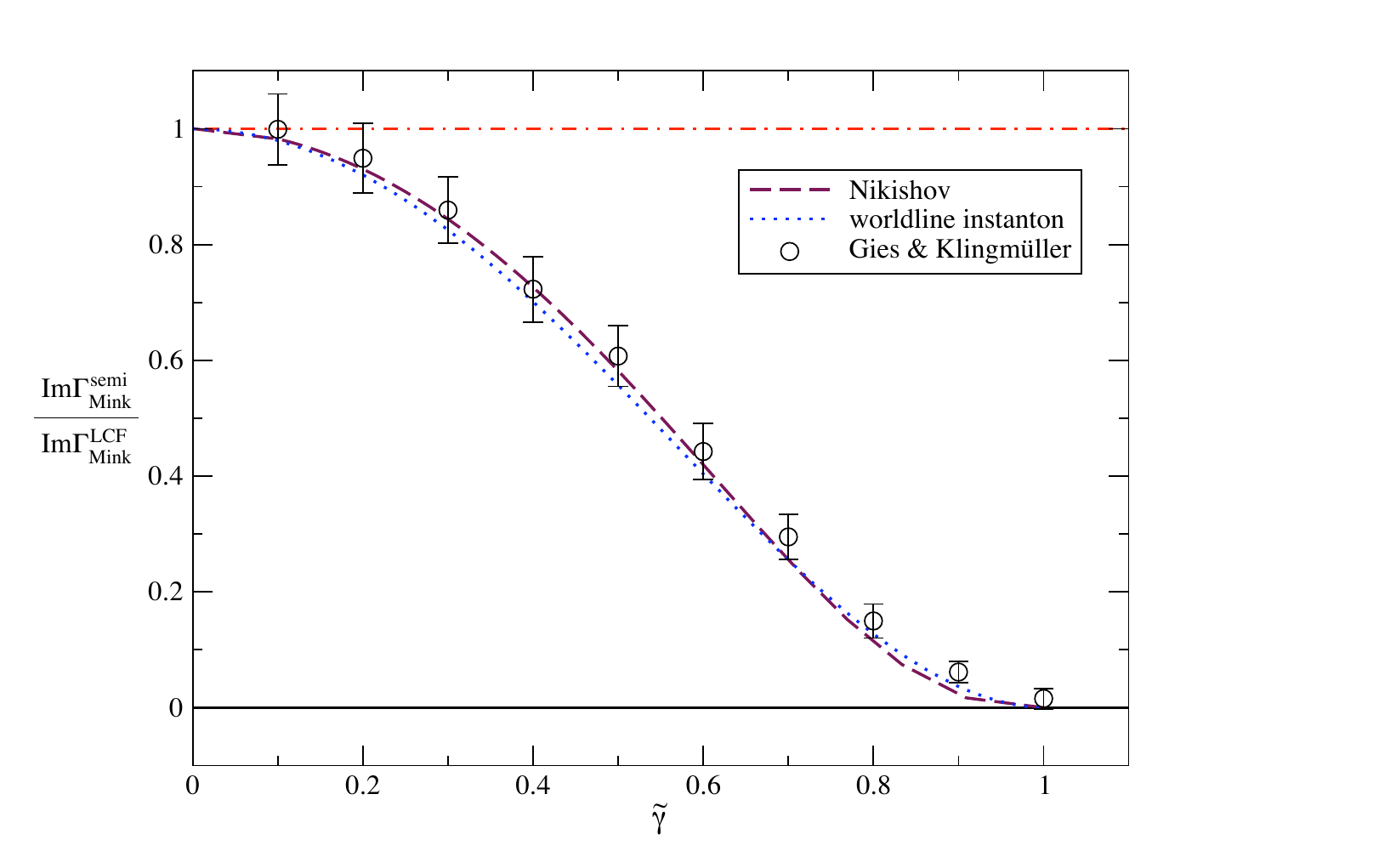}}
\label{singlebumpcompare}
\caption{Plot of the imaginary part of the effective action 
for the field $E(x)=E{\rm sech}^2(kx)$ as
a function of the inhomogeneity parameter $\tilde\gamma$,
normalized by the weak field limit of the
 ``locally constant field approximation'' \cite{wlinst2}.
The dotted line shows the result obtained from the
worldline instanton approximation. The dashed line is the same
ratio using a numerical integration of Nikishov's exact expression
\cite{nikishov}. The circles represent the worldline Monte Carlo
results of \cite{giekli}, evaluated for $eE/m^2 =1$.}  
\end{figure}

The single-bump field also provides an excellent opportunity to test the validity of the
semiclassical approximation, since for this case an exact integral representation
has been obtained by Nikishov \cite{nikishov}, suitable for numerical evaluation,
and moreover there is a worldline Monte Carlo result \cite{giekli}. 
As shown in figure \ref{singlebumpcompare}, all three results are in close agreement
over the whole range of the inhomogeneity parameter $\tilde\gamma$. 

Note that the
imaginary part vanishes for $\tilde\gamma > 1$, which in the instanton approach
simply means that instanton solutions cease to exist. Although mathematically this 
absence of instanton solutions allows one
to conclude the vanishing of the imaginary part only in the semiclassical approximation,
this example of the single-bump case supports the conjecture that
it might, in fact, signal the complete absence of pair production.

Although the method works very well for these classical cases, these are
rather special configurations, and known to be amenable also to
WKB methods. The existence of closed-form expressions
for the worldline instanton can be expected only for a very restricted
classes of fields. Much more interesting is the fact that the worldline instanton
equations (\ref{Lorentz}) and the zero mode equations 
(\ref{Lambdazeromodes}) are ordinary differential equations, which leads us to
expect that they can be solved numerically for more general classes of fields
than have been treated by WKB. Dunne and Wang \cite{dunwan} have very recently
applied this numerical approach to a class of electric fields 
which depend nontrivially on two spatial coordinates, 
parametrized by

\vspace{-10pt}

\begin{figure}[h]
\centerline{\includegraphics[height=.3\textheight]{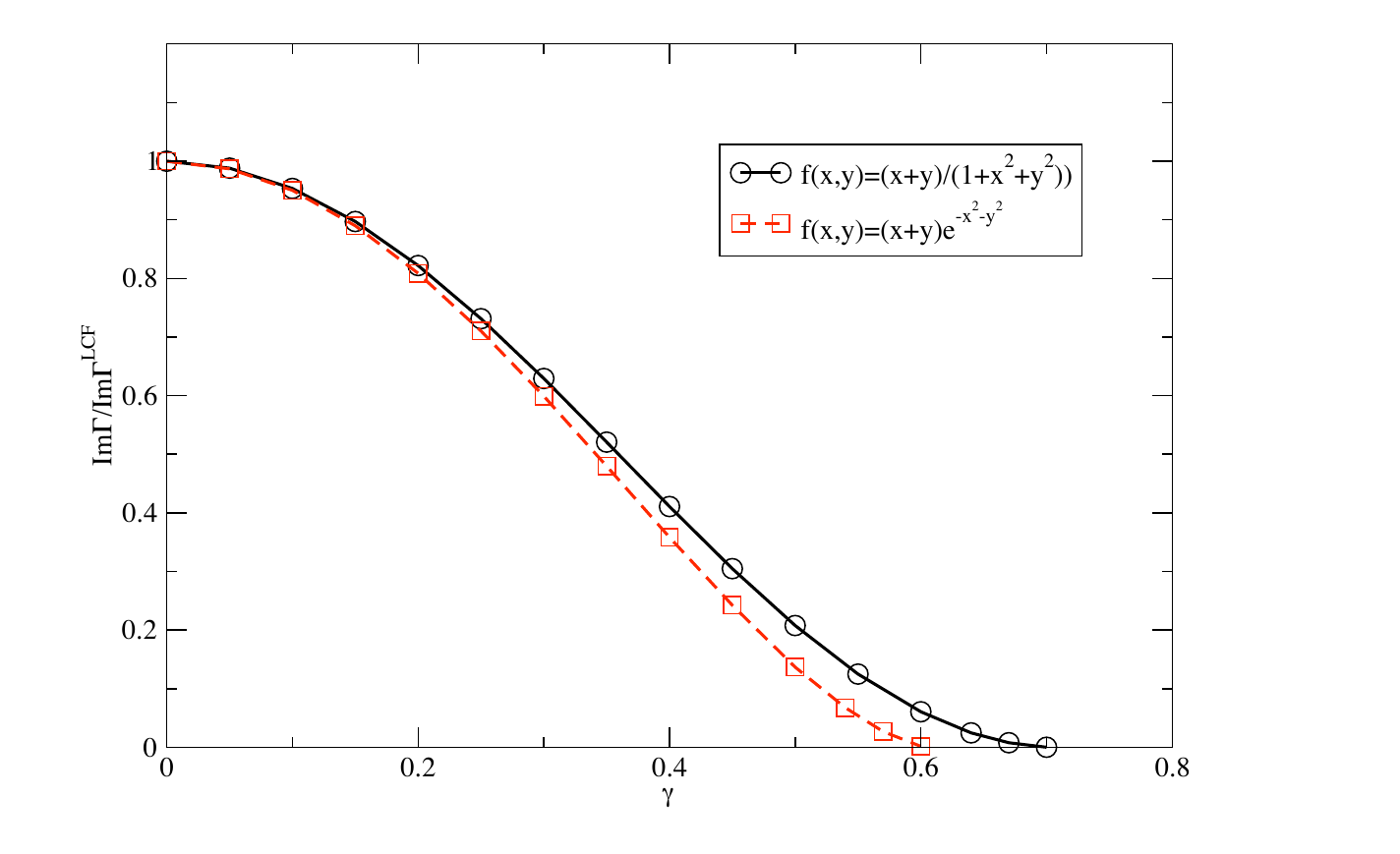}}
\caption{Pair creation rates for two nonplanar cases.}
\label{dunnwang}
  \end{figure}

\bear
A_4({\vec x}) = -i {E\over k}\,f(\vec x)
\label{gauge}
\ear
($ \gamma = mk / eE$).
Fig. \ref{dunnwang}
\footnote{Reproduced from \cite{dunwan} with the permission of the authors.}
shows their results for two examples, 

\bear
f(\vec x) &=& {k(x_1+x_2)\over 1+k^2( x_1^2+x_2^2)}\non\\
f(\vec x) &=& k(x_1+x_2)\,\e^{-k^2(x_1^2+x_2^2)}\non\\
\label{twoexamples}
\ear

\no
Note that again ${\rm Im} \Gamma$ vanishes for large inhomogeneities, i.e.
for fields of insufficient extent.

Finally, making the transition from scalar to spinor QED in the instanton
approach is straightforward if we use Feynman's original implementation of
spin in the worldline path integral. Up to the global normalization,
it amounts to multiplying with the spin factor (\ref{defspinfactor}) 
evaluated on the instanton trajectory, $S[x^{\rm cl},A]$.
For the classical cases discussed above, and
more generally for all ``planar'' fields the path ordering in the
spin factor has no effect.  Surprisingly, one obtains simply
\cite{wlinst1}


\bear
S[x^{\rm cl},A] &=& 4\cos \biggl[
eT \int_0^1 du\, E(x(u)) \biggr]
= 4(-1)^n
\label{spinfactorplanar}
\ear
with $n$ the winding number. Thus for planar fields
one finds the same simple relation between 
$ {\rm Im} {\cal L}_{\rm scalar}(E)$ and 
$ {\rm Im} {\cal L}_{\rm spinor}(E)$
as in the constant $ E$ case, eqs.(\ref{ImSpin}),(\ref{ImScal})!
It is an interesting open question whether this property might even
extend to general electric fields.

\vspace{-10pt}

\section{5. Summary}

The purpose of this 
short review was to show that the worldline approach in its various versions is turning
into an efficient alternative to second-quantized methods for an increasing range of problems in
QED. Unfortunately, despite of the restriction to QED 
it was not possible here to give due credit to all
relevant work. Among other things, it was not possible here to discuss the worldline variational
approach of \cite{rossch,alrosc}, nor the nonperturbative propagator calculations of
\cite{satjgr}.

Let me conclude with summarizing the main advantages which one can hope
to achieve using the worldline formalism:
(i) Compact parameter integral representations for arbitrary QED multiloop amplitudes
(ii) Easy implementation of constant external fields
(iii) Reliable numerical (Monte Carlo) results for one loop 
effective actions and Casimir energies 
(iv) Reliable numerical results for pair creation rates in arbitrary electric fields.



\end{document}